\documentstyle[twocolumn,floats,aps]{revtex} 

\input epsf         % defines \epsfbox and supporting macros
\epsfverbosetrue    % messages will show height and width
\def\postscript#1{\begin{center}\leavevmode
\hbox{\epsfxsize=0.95\columnwidth\epsfbox{#1}}\end{center}}

\begin{document}

%%%%%%%%%%%%%%%%%%%%%
\twocolumn[\hsize\textwidth\columnwidth\hsize\csname@twocolumnfalse%
\endcsname

\draft

\title{Normal-state magnetic susceptibility in a bilayer cuprate}

\author{W. C. Wu,$^1$ B. W. Statt,$^2$ Ya-Wei Hsueh,$^2$ and J. P. Carbotte$^1$}
\address{$^1$Department of Physics and Astronomy, McMaster University,
Hamilton, Ontario, Canada L8S 4M1}
\address{$^2$Department of Physics, University of Toronto,
Toronto, Ontario, Canada M5S 1A7}
 
\date{\today}

\maketitle

\begin{abstract}
The magnetic susceptibility of high-$T_c$ superconductors
is investigated in the normal state using a coupled bilayer model.
While this model describes in a natural way the
normal-state pseudogaps seen in $c$-axis optical conductivity
on underdoped samples, it predicts a weakly increasing susceptibility
with decreasing temperature and cannot explain the magnetic pseudogaps 
exhibited in NMR measurements. Our result, together with some experimental 
evidence suggest that the mechanism governing
the $c$-axis optical pseudogap is different from that for
the $a$-$b$ plane magnetic pseudogap. 
\end{abstract}

%%%%%%%%%%%%%%%%%%%
]

\vskip 0.1 true in
\narrowtext

Normal-state pseudogaps of underdoped samples
have been one of the important subjects in the 
study of high-$T_c$ cuprates in the past few years \cite{Levi}.
Experimentally strong decrease of NMR Knight shift and spin-lattice
relaxation rate with decreasing temperature
has been observed in various normal-state underdoped samples 
below some cross over temperature $T^*$ 
\cite{Takigawa,Walstedt92,Bankay,Stern,Hsueh97}.
These decreases have often been considered in terms of the opening of a
``spin gap'' due to the antiferromagnetic (AF)
spin fluctuations \cite{MMP,MPT,CPS} or, alternatively,
the notion of ``preformed Cooper pairs''
in the case of significant order parameter phase fluctuations \cite{EK95}.
Measurements on dc resistivity for underdoped samples
also reveal that the in-plane resistivity $\rho_{ab}(T)$ 
deviates from a linear in temperature behavior 
at some temperature consistent with the crossover temperatures
$T^*$ in NMR experiments \cite{TMTU}. 
On the other hand, the $c$-axis optical conductivity $\sigma_c(\omega)$ 
\cite{Homes93,Homes95,Basov96,Reedyk97} 
exhibits a striking gap-like depression at low frequency
and the $c$-axis resistivity $\rho_c(T)$ displays a upturn semiconducting
feature  \cite{TMTU} in the normal-state underdoped samples,
characteristic of the formation of some kinds of gap.

A central question in the study of pseudogaps is --
are the pseudogaps seen in the $c$-axis optical conductivity
related {\em at all} \ to the magnetic pseudogaps
seen in NMR experiments? Or, 
could there be two different mechanisms?
While there is currently no consensus on this point, we begin by comparing 
first some of the properties of these two pseudogaps seen in experiments.
Consider the magnitude and anisotropy of these two pseudogaps:
(i){\em ~Magnitude} -- Although one does not measure the 
magnetic pseudogap directly, an analysis done by Williams {\em et al.}
\cite{Williams} of Knight-shift data based on quasiparticle excitation
spectrum given by
$E_{\bf k}=[\epsilon_{\bf k}^2+\Delta_{\bf k}^2]^{1/2}$
where $\Delta_{\bf k}$ is the normal-state pseudogap
shows that the magnitude of magnetic pseudogap is strongly 
dependent on doping or carrier density
(larger gap with lower doping) and scales with
the crossover temperature $T^*$.
In contrast, the $c$-axis optical conductivity 
measured by Homes {\em et al.} \cite{Homes95} on YB$_2$C$_3$O$_{6+x}$ 
(YBCO) at different doping $x$ indicates that
the frequency ranges where conductivity is suppressed 
(direct measurement for the size of the gap) are almost independent of doping.
Moreover, a comparison \cite{Hsueh97}
of YBCO and Pb$_2$Sr$_2$(Y,Ca)Cu$_3$O$_{8+\delta}$ (PSYCCO)
data gives that the optical pseudogap does not scale with $T^*$. 
(ii) {\em Anisotropy} --
NMR \cite{Williams} experiments reveal that
the magnetic pseudogap has $d$-wave like anisotropy
(consistent with the angular resolved photoemission spectroscopy (ARPES) 
\cite{Loeser96} result), while  
the {\em flat} spectral weight seen within the pseudogap region
in $c$-axis optical data implies that  the optical
pseudogap is isotropic ($s$-wave like).
The above comparison seems to suggest that the
magnetic pseudogap is intimately connected to the superconducting gap 
and reflects the pairing mechanism,
while the optical pseudogap is due to a different mechanism.

Recently Atkinson and Carbotte \cite{AC97-1} and with
Wu \cite{AWC97,WAC97-1} have used a coupled bilayer model
to calculate various $c$-axis properties of the high $T_c$ cuprates.
Within this model, a minimum energy difference $\Delta$
between two non-degenerate bands 
(associated with two layers coupled by a single-particle
hopping $t_\perp$) is introduced. This band gap $\Delta$
-- turns out to be the pseudogap seen
in the $c$-axis optical conductivity -- and gives the
suppression of the joint density of states in the interband transitions. 
The key feature of this coupled bilayer model is that
the $c$-axis optical conductivity can be generally separated into  
intraband (proportional to $t_\perp^4$) and interband 
(proportional to $t_\perp^2$) parts. Therefore, when $t_\perp$ is small 
which is the condition appropriate to the underdoped regime
\cite{ZCP}, interband contribution dominates over the usual intraband 
contribution and, as a result, the pseudogap is visible
(non-Drude like behavior).
In contrast in the larger $t_\perp$ 
optimally-doped or overdoped regime, intraband dominates
and one retains the more usual no-pseudogap Drude-like behavior. 

\begin{figure}[tb]
\vspace{-0.8cm}
\postscript{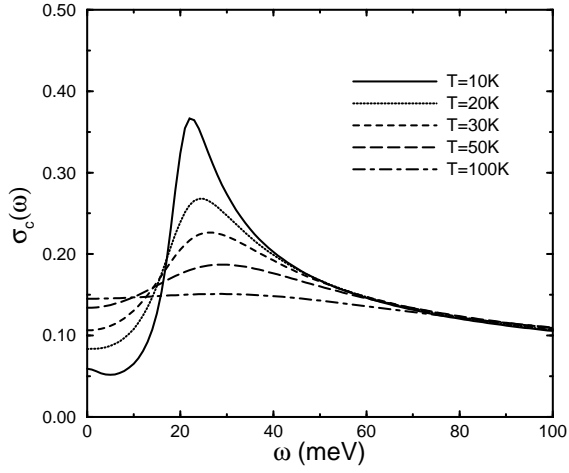}
\vspace{-0.5cm}
\caption{Normal-state $c$-axis conductivity as a function of photon
energy for a plane-chain bilayer at different
temperatures with a small $t_\perp=2{\rm meV}$.
The minimum energy difference between the two bands for
a given ${\bf k}$ in the first 2D Brillouin zone is $\Delta=20{\rm meV}$
(taken from Fig.~1 in Ref.~\protect\cite{WAC97-1}).}
\label{fig1}
\end{figure}

It is shown in Refs.~\cite{AC97-1,WAC97-1} that
the coupled bilayer model explains in a natural way
the pseudogap seen in the $c$-axis optical
conductivity for underdoped, and gives an explanation of
why it is not seen in optimally-doped or overdoped samples
(see Fig.~\ref{fig1}).
Both temperature and frequency dependence of the entire
spectrum seen in experiments \cite{Homes93,Homes95}
are well described by this model.
It is thus not surprisingly that
the semiconductor-like upturn feature at low temperature seen
in the $c$-axis dc resistivity is also understood in this model
\cite{WAC97-1}.  Since this model emphasizes the band structure correlation
arising from two non-degenerate coupled layers,
a simple test can be immediately done for the
single-layer La$_{2-x}$Sr$_x$CuO$_4$ (LSCO) which
is found by Basov {\em et al.} \cite{Basov95}
to exhibit no pseudogap in the $c$-axis optical conductivity
or a very weak pseudogap with large energy scale $\sim 0.1{\rm eV}$
by Uchida {\em et al.} \cite{Uchida}.
The latter case accompanied with a 
semiconducting upturn feature of $\rho_c(T)$  \cite{Boebinger}
strongly suggests that while there is only one conducting layer, 
in practice there are {\em two} bands separated by a
large energy difference $\Delta\sim 0.1{\rm eV}$ in LSCO.
We note that in this case Startseva {\em et al.} \cite{Startseva}
find a strong in-plane pseudogap with smaller energy scale
indicating no correlation between in-plane and out-of-plane energy scale
for the pseudogap. 
 
In the following, we use the bilayer model to 
calculate the magnetic susceptibility.
We consider the spin raising ($+$)
and lowering ($-$) operators for layer $i$ ($i$=1,2) defined by
$\sigma^+_{i,{\bf q}}$=$\sum_{\bf k}c^\dagger_{i,{\bf k+q},\uparrow} 
c_{i,{\bf k},\downarrow}$ and 
$~\sigma^-_{i,{\bf q}}$=$\sum_{\bf k}c^\dagger_{i,{\bf k+q},\downarrow} 
c_{i,{\bf k},\uparrow}$. The {\em transverse} (parallel to $a$-$b$ plane)
magnetic susceptibility $\chi_{ij}$ which we shall concentrate 
on, is then given by \cite{DS}

\begin{equation}
\chi_{ij}({\bf q},i\omega_n)=2\mu_e^2 
\int_0^{\beta}d\tau e^{i\omega_n \tau}
\langle{\rm T}_\tau \sigma_{i,{\bf q}}^- (\tau)
\sigma_{j,-{\bf q}}^+ (0)\rangle,
\label{eq:chi.spin}
\end{equation}
where $\mu_e$ is the Bohr magneton,
$\chi_{11},\chi_{22}$ denote the 
in-plane spin correlation, and
$\chi_{12},\chi_{21}$ denote the out-of-plane (or cross) spin correlation.
Due to its local nature, an NMR experiment is able to study 
the spin susceptibility $\chi_{ij}({\bf q},\omega)$ individually.
For example, the Knight shift ($K_s$) for an atom in layer $i$
is directly proportional to the real part of the planar spin susceptibility, 
$K_s(T)\sim \chi_{ii}^\prime({\bf q}=0,\omega=0)$
and the spin-lattice relaxation rate (per temperature unit)
is related to imaginary-part susceptibility over a weighted sum,
$(1/T_1 T)_{i} \sim \sum_{\bf q}|A({\bf q})|^2
\chi_{ii}^{\prime\prime}({\bf q},\omega\simeq 0)/\omega$.
While, the information contained in the cross spin susceptibility $\chi_{12}$
can be probed by a spin echo double resonance (SEDOR) experiment
(see, for example, Ref.~\cite{Slichter_book}).

In terms of a free electron system, the susceptibility
(\ref{eq:chi.spin}) can be reduced to
\begin{eqnarray}
&&\chi_{ij}({\bf q},i\omega_n)=-2\mu_e^2
\sum_{\bf k}{1\over \beta}\sum_{\nu_n}\nonumber\\
&&~~~{\rm Tr}[G_0({\bf k},i\nu_n)\gamma_i
G_0({\bf k+q},i\nu_n - i\omega_n)\gamma_j],
\label{eq:various.res.fcns}
\end{eqnarray}
tracing over the two bands, where 
$\gamma_1$={\scriptsize 
$\left[
\begin{array}{cc}
1 & 0 \\ 
0 & 0 \\
\end{array}
\right]$} and 
$\gamma_2$={\scriptsize
$\left[
\begin{array}{cc}
0 & 0 \\
0 & 1 \\
\end{array}
\right]$}
are the vertices and

\begin{eqnarray}
G_0^{-1}({\bf k},i\nu_n)=
\left[
\begin{array}{cc}
i\nu_n-\xi_1({\bf k}) & -t(k_z) \\ 
-t(k_z) & i\nu_n-\xi_2({\bf k})\\
\end{array}
\right]
\label{eq:G0}
\end{eqnarray}
is the Green's function matrix.
The parameter $\xi_i$ is the band structure for isolated layer $i$ and
$t$ is the coupling between the two layers.
Considering only the ${\bf q}\rightarrow 0$ case,
the trace operator in (\ref{eq:various.res.fcns})
enables one to calculate $\chi_{ij}$ 
in a convenient frame in which the Green's function matrix is diagonal. 
Consequently,

\begin{eqnarray}
&&\chi_{ij}({\bf q}=0,i\omega_n)=-2\mu_e^2
\sum_{\bf k}{1\over \beta}\sum_{\nu_n} \nonumber\\
&&~~~ {\rm Tr}[\hat{G}_{0}({\bf k},i\nu_n)\hat{\gamma}_i
\hat{G}_{0}({\bf k},i\nu_n - i\omega_n)\hat{\gamma}_j],
\label{eq:various.res.fcns.hat}
\end{eqnarray}
where 
 
\begin{eqnarray}
&&~~~\hat{G}_{0}^{-1}({\bf k},i\nu_n)=
\nonumber\\
&&\left[
\begin{array}{cc}
i\nu_n-\epsilon_{+}+i\Gamma_+ {\rm sgn}(\nu_n) & 0 \\ 
0 & i\nu_n-\epsilon_{-}+i\Gamma_- {\rm sgn}(\nu_n) \\
\end{array}
\right],
\label{eq:G0.hat}
\end{eqnarray}
with $\Gamma_\pm$ introduced as the total scattering rate in band $\pm$.
The two renormalized bands are

\begin{equation}
\epsilon_\pm={\xi_1+\xi_2\over 2}\pm
\sqrt{\left({\xi_1-\xi_2\over 2}\right)^2+|t|^2}
\label{eq:spectrum2}
\end{equation}
and the rotated vertices are

\begin{eqnarray}
\hat{\gamma}_1 = 
\left[
\begin{array}{cc}
\alpha_{11} &\alpha_{12}  \\ 
\alpha_{12} & \alpha_{22} \\
\end{array}
\right]~~;~~
\hat{\gamma}_2 = 
\left[
\begin{array}{cc}
\alpha_{22}&-\alpha_{12}\\
-\alpha_{12}&\alpha_{11}\\
\end{array}
\right],
\label{eq:diag.gamma_12}
\end{eqnarray}
where $\alpha_{11}\equiv(\xi_1-\epsilon_-)/(\epsilon_+ -\epsilon_-)$,
$\alpha_{22}\equiv(\epsilon_+ -\xi_1)/(\epsilon_+ -\epsilon_-)$, and
$\alpha_{12}\equiv |t|/(\epsilon_+ -\epsilon_-)$.
Substituting (\ref{eq:G0.hat}) and (\ref{eq:diag.gamma_12}) into
(\ref{eq:various.res.fcns.hat}), we obtain
 
\begin{eqnarray}
\chi_{11}(i\omega_n)&=&-2\mu_e^2
\sum_{\bf k}{1\over \beta}\sum_{\nu_n}\left[
\Bigl(\alpha_{11}^2 G_{11}G_{11}^\prime+\alpha_{22}^2
G_{22}G_{22}^\prime \Bigr)\right.\nonumber\\
&+&\left. \alpha_{12}^2\Bigl(G_{11}G_{22}^\prime+ G_{22}G_{11}^\prime 
\Bigr)\right],
\nonumber\\
\chi_{22}(i\omega_n)&=&-2\mu_e^2
\sum_{\bf k}{1\over \beta}\sum_{\nu_n}\left[
\Bigl(\alpha_{22}^2 G_{11}G_{11}^\prime+\alpha_{11}^2
G_{22}G_{22}^\prime \Bigr)\right.\nonumber\\
&+& \left.
\alpha_{12}^2\Bigl(G_{11}G_{22}^\prime+ G_{22}G_{11}^\prime \Bigr)\right],
\nonumber\\
\chi_{12}(i\omega_n)&=&\chi_{21}(i\omega_n)=-2\mu_e^2
\sum_{\bf k}{1\over \beta}\sum_{\nu_n}\left[
\alpha_{11}\alpha_{22}\Bigl(G_{11}G_{11}^\prime\right.\nonumber\\
&+&\left. G_{22}G_{22}^\prime \Bigr)-
\alpha_{12}^2\Bigl(G_{11}G_{22}^\prime+ G_{22}G_{11}^\prime \Bigr)\right],
\label{eq:chis}
\end{eqnarray}
where $G_{ii}\equiv G_{ii}({\bf k},i\nu_n)$ and
$G_{ii}^\prime\equiv G_{ii}({\bf k},i\nu_n - i\omega_{n})$
are the elements of (diagonal) Green's function matrix in (\ref{eq:G0.hat}).
Clearly for each $\chi_{ij}$ in (\ref{eq:chis}), 
the first term corresponds
to an {\em intraband} contribution and the second term corresponds
to an {\em interband} contribution.
One sees that the in-plane $\chi_{ii}$ and cross $\chi_{ij}~(i\neq j)$
have equal but opposite-sign interband contributions.

The frequency sum in (\ref{eq:chis})
can easily be transformed into an integral such as
(using (\ref{eq:G0.hat}))

\begin{eqnarray}
&&\left[{1\over \beta}\sum_{\nu_n} G_{ii}({\bf k},i\nu_n)
G_{jj}({\bf k},i\nu_n - i\omega_{n})\right]^\prime\nonumber\\
&=&\int_{-\infty}^\infty
{dx\over 2\pi}~{2\Gamma\over [(x-\epsilon_i)^2+\Gamma^2]
[(x-\omega-\epsilon_j)^2+\Gamma^2]}\nonumber\\
&\times&\Bigl[f(x)(x-\omega-\epsilon_j)+f(x-\omega)(x-\epsilon_i)\Bigr],
\label{eq:A.int}
\end{eqnarray}
where $f(x)$ is the Fermi distribution function and
we have redefined $\epsilon_1\equiv\epsilon_+$ and
$\epsilon_2\equiv\epsilon_-$. The scattering rates are
simply assumed to be identical for both bands,
$\Gamma_+=\Gamma_-\equiv\Gamma$. To obtain analytical results, 
we consider the two isolated layers to have
the following simple band structures

\begin{eqnarray}
\xi_1(k_x,k_y)&=&{\hbar^2\over 2m}k_\parallel^2-\mu+\Delta
\nonumber\\
\xi_2(k_x,k_y)&=&{\hbar^2 \over 2m}k_\parallel^2-\mu,
\label{eq:simpleband.pp}
\end{eqnarray}
where $k_\parallel\equiv (k_x^2 +k_y^2)^{1/2}$ is the 2D
layer momentum, $\mu$ is the chemical potential,
and $\Delta$ corresponds to the energy difference between these 
two layer bands (in general,
$\Delta$ corresponds to a {\em minimum}
energy difference between two bands).
For YBCO, we consider the CuO$_2$ plane as layer 1 which has generally 
more (hole) carriers (or less electrons) than the CuO chain which is 
layer 2.  While the band structure (\ref{eq:simpleband.pp}) is 
oversimplified, it has captured the essential physics built in this 
coupled bilayer model.  Taking into account
the periodicity of crystals, the layer-layer coupling 
is chosen to be $t(k_z)=2t_\perp \cos(k_z d/2)$, with the parameters 
$t_\perp$ the coupling strength and $d$ the spacing between two layers.
When $t_\perp$ is small ($|t|\ll \Delta$), one may expand
$\alpha_{11}\simeq 1-t^2/\Delta^2$, $\alpha_{22}\simeq t^2/\Delta^2$,
and $\alpha_{12}\simeq |t|/\Delta$.
This in turn implies that
for the planar spin susceptibilities $\chi_{11}$ and $\chi_{22}$,
the intraband term is of order 1 and
the interband term is proportional to
$t_\perp^2$, while $\chi_{12}$ has 
both intraband and interband terms proportional to $t_\perp^2$.

\begin{figure}[tb]
\vspace{-0.8cm}
\postscript{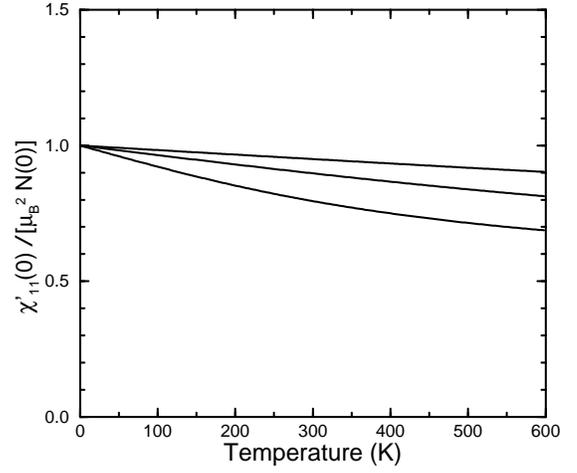}
\vspace{-0.5cm}
\caption{Temperature dependence of the theoretical planar magnetic
susceptibility $\chi_{11}^\prime(0)$ given in (\protect\ref{eq:Ks}).
The curves from top to bottom correspond to
$\mu=400$, $200$, and $100{\rm meV}$
with fixed $\Delta=20{\rm meV}$, $t_\perp=3{\rm meV}$,
and $\Gamma(T)=20{\rm meV}$ at $100K$ and is linear in temperature.}
\label{fig2}
\end{figure}

With the above simplifications, 
the real part of the magnetic susceptibilities in (\ref{eq:chis})
at zero frequency can be found to be
(keeping only second order terms in $t_\perp$)

\begin{eqnarray}
\chi_{11}^\prime(0)&=&\mu_e^2 N(0) \biggl[
{1\over 2}+{1\over \pi}{\rm arctan}\left({\mu-\Delta\over\Gamma}\right)
+{4\over \pi}{t_\perp^2\over \Delta^2}(\beta-\gamma) \biggr]
\nonumber\\
\chi_{22}^\prime(0)&=&\mu_e^2 N(0) \biggl[
{1\over 2}+{1\over \pi}{\rm arctan}\left({\mu\over\Gamma}\right)\nonumber\\
&+&{4\over \pi}{t_\perp^2\over \Delta^2}\left({\mu-\Delta\over \mu}\beta-
\gamma\right) \biggr]
\nonumber\\
\chi_{12}^\prime(0)&=&\chi_{21}^\prime(0)=-{2\over \pi}
{t_\perp^2\over \Delta^2} \mu_e^2 N(0) \biggl[
{2\mu-\Delta\over \mu}\beta-2\gamma \biggr]
\label{eq:Ks}
\end{eqnarray}
where $N(0)=m/\pi \hbar^2$ is the 2D (constant)
density of states for free electron gas and we have defined

\begin{eqnarray}
\beta(T)&=&{\mu\over\Delta}\left[{\rm arctan}\left({\mu\over \Gamma}\right)
-{\rm arctan}\left({\mu-\Delta\over\Gamma}\right)\right]
\nonumber\\
\gamma(T)&=&{\Gamma\over 2\Delta}{\rm ln}\left[{\mu^2+\Gamma^2\over
(\mu-\Delta)^2+\Gamma^2}\right].
\label{eq:beta.gamma}
\end{eqnarray}
In deriving (\ref{eq:Ks}), we have made use of the fact that
$-\partial f(x)/\partial x \approx \delta(x)$, which is appropriate
for not too high temperatures.
In Fig.~\ref{fig2}, we plot the temperature-dependent
layer-1 Knight shift (proportional to $\chi_{11}^\prime(0)$ in
(\ref{eq:Ks})) using the parameters which is appropriate to and has 
successfully described the $c$-axis optical pseudogap for 
underdoped YBa$_2$Cu$_3$O$_{6.6}$:
$\mu=100$ to $400{\rm meV}$, $\Delta=20{\rm meV}$, $t_\perp=3{\rm meV}$,
and $\Gamma(T)=20{\rm meV}$ at $100K$ and is linear in temperature.
It is clearly shown in Fig.~\ref{fig2}
that the Knight shift increases only slightly as temperature 
decreases and does not exhibit
the decreasing feature characterizing the
magnetic pseudogap seen in experiments.
Since in YBCO, $\Delta\ll\mu$, $\chi^\prime_{22}(0)$ is found to be 
similar to $\chi_{11}^\prime(0)$ of Fig.~\ref{fig2}.
Apart from the magnitude of $t_\perp$,
it can be shown quite generally that 
when $\Gamma(T)<\mu$, $~\beta-\gamma\sim -\Gamma^5/[\mu^3(\mu^2+\Gamma^2)]$ 
which is small.  This means that the last term
in $\chi_{11}^\prime(0)$ or $\chi_{22}^\prime(0)$ in
(\ref{eq:Ks}) does not contribute much and
one expects a similar temperature-dependent Knight shift
as given in Fig.~\ref{fig2}, even for overdoped 
YBa$_2$Cu$_3$O$_7$ for which $t_\perp$ is much larger.
Naively if one applies the result $\chi_{11}^\prime(0)$ in (\ref{eq:Ks})
to both YBCO with smaller $\Delta=20{\rm meV}$
and LSCO with larger $\Delta\sim 0.1{\rm eV}$
and assumes $\mu$ and $\Gamma$ identical for both cases,
one expects more significant increasing Knight shifts with 
decreasing temperature in overdoped LSCO compared
to those in overdoped YBCO which
seems consistent with experiments \cite{Nakano94}.

In contrast to the $c$-axis conductivity where
the interband term ($\propto t_\perp^2$) dominates over 
the intraband term ($\propto t_\perp^4$) in the small $t_\perp$ underdoped
case, for the planar spin susceptibility, 
the interband term ($\propto t_\perp^2$) 
is almost canceled by an equal contribution arising from
intraband terms. Consequently, the planar spin susceptibility
is {\em not} very much dependent on the interband transition (or, equivalently,
the magnitude of $t_\perp$), 
and, as a result, also not very dependent on temperature.
Thus the Knight shift derived from the simple coupled bilayer model
has only a {\em weak} temperature dependence mainly
due to impurity scattering effect (i.e., $\Gamma(T)$).

In principle one can measure directly the effect of $t_{\perp}$ with
SEDOR \cite{Slichter_book}. An effective
coupling $a$ exists between spins on the two layers \
$a {\bf I_1}\cdot{\bf I_2}$. In terms of our model 
$a \sim \hbar A^2 \chi_{12}/\mu_B^2$ where $A$ is the
hyperfine coupling constant. In order to obtain an estimate 
for $a$ we take $A$ to be the isotropic transferred hyperfine constant 
for Cu atoms in the CuO$_2$ planes of YBCO where $A/^{63}\gamma$ = 
82 kOe and take $N(0)$ = 2 states/eV--Cu.
Evaluating $\chi_{12}$ at $T =$ 300 K with $\mu = $ 100 meV yields
$\chi_{12}^{\prime}/\mu_B^2N(0)$ = 0.0016. Thus the effective coupling rate
is $a \simeq $ 15 s$^{-1}$ which is about 3 orders of magnitude too 
small to observe experimentally \cite{Statt}.
 
While the theory given in Eq.~(\ref{eq:Ks}) does not
describe the magnetic pseudogaps for underdoped samples,
several predictions and conclusions can be drawn in 
connection with experimental data.
First, there exists a mechanism (not the interband effect)
which operates most profoundly in the underdoped samples
and leads to the suppression of Knight shifts (magnetic pseudogaps).
Secondly, this mechanism probably
contributes to the pseudogap effects seen
in {\em all} the $a$-$b$ plane measurements, including the charge-associated
in-plane dc resistivity $\rho_{ab}(T)$, ARPES, in-plane scattering rate
$1/\tau(\omega,T)$ in the infrared reflectivity, and the spin-associated
NMR measurements and is of only secondary importance in $c$-axis transport.  
Thirdly, the interband effect we propose should   
dominate in the $c$-axis transport and
is strongly associated with the
pseudogap effects seen in $c$-axis optical and dc 
resistivity measurements.
Fourth, as indicated in experiment on YBa$_2$Cu$_4$O$_8$
\cite{Bankay}, the magnetic pseudogap mechanism
has a much stronger effect on the CuO$_2$ plane (deeper normal-state
Knight-shift drop) than CuO chain (weaker or almost no drop).

In summary, while at present one cannot rule out
that a single mechanism may govern both the $c$-axis optical and $a$-$b$ plane 
magnetic pseudogaps, we have shown through our
theoretical studies as
well as with experimental data that a two-mechanism picture is favored.

The authors thank Bo\v{z}idar Mitrovi\'{c}, 
Tom Timusk, and Tatiana Startseva for very stimulating discussion.
This work was supported by Natural Sciences and Engineering Research Council
(NSERC) of Canada and Canadian Institute for Advanced Research (CIAR).

\end{document}